\documentclass[a4paper,12pt]{article}
\usepackage{amsmath, amsthm}
\usepackage{amssymb,amsfonts}
\usepackage{amsmath,amssymb}
\usepackage[mathscr]{euscript}
\usepackage{bm}
\usepackage{enumerate}
\usepackage{csquotes}
\usepackage{fancyhdr}
\usepackage{graphicx}
\usepackage{hyperref}
\usepackage{authblk}

\usepackage{amssymb}

\theoremstyle{definition}
\newtheorem{definition}{Definition}[section]

\newtheorem{remark}[definition]{Remark}

\theoremstyle{plain}
\newtheorem{theorem}[definition]{Theorem}
\newtheorem{lemma}[definition]{Lemma}

\numberwithin{equation}{section}

\usepackage{color}   %May be necessary if you want to color links
\usepackage{hyperref}
\usepackage{url}
\begin{document}
	\date{}
	\author[1]{Wajid M. Shaikh}
	\author[2]{Rupali S. Jain}
	\author[3]{B. Surendranath Reddy}
	\author[4]{Bhagyashri S. Patil}
	\author[5]{Sahar M. A. Maqbol}
	\affil[1,4]{Research Scholar, School of Mathematical Sciences, SRTMU Nanded}
	\affil[2]{Associate Professor, School of Mathematical Sciences, SRTMU Nanded}
	\affil[3]{Assistant Professor, School of Mathematical Sciences, SRTMU Nanded}
	\affil[5]{Department of Mathematics, Hodeidah University, P.O. 3114, Al-Hodeidah, Yemen.}
	\title{
		{A new approach to construct minimal linear codes over $\mathbb{F}_{3}$  }\\
	}
	\maketitle
	\begin{abstract}
		
	In this article, we present  two new approaches to construct minimal linear codes of dimension $n+1$ over $\mathbb{F}_{3}$ using  characteristic and ternary functions. We also obtain the weight distributions of these constructed minimal linear codes. We further show that a specific class of these codes violates Ashikhmin-Barg condition.

	\end{abstract}
	\section{Introduction}

Linear codes are fundamental in contemporary communication systems offering vital error control capabilities, crucial for ensuring dependable data transmission across a spectrum of applications from telecommunications to digital broadcasting. Among these, minimal linear codes stand out as a distinct class, finding significant utility in secret sharing, data storage, and information processing systems within modern technology landscapes \cite{4},\cite{8}. As such, the development and analysis of minimal linear codes have been subject to extensive research efforts by scholars  \cite{5},\cite{6},\cite{7}.
\par In 1998,  Ashikhmin and Barg\cite{1} gave a sufficient condition for linear code to be minimal, which is stated below.
\begin{lemma}\cite{1}\label{lemma0.1}[Ashikhmin-Barg]
	Let $p$ be a prime, then a linear code $\mathscr{C}$ over $\mathbb{F}_{p}$ is minimal if
	\begin{equation*}
	\frac{{wt}_{min}}{{wt}_{max}}>\frac{p-1}{p}
	\end{equation*}
where, ${wt}_{min}$ and ${wt}_{max}$ denote the minimum and maximum non-zero Hamming weights for $\mathscr{C}$, respectively.
	\end{lemma}
\par Many minimal linear codes can be constructed based on the lemma mentioned above, however, the construction of minimal linear codes that violate the Ashikhmin-Barg condition presents an intriguing research avenue. In the year 2018, Chang and Hyun\cite{2} constructed an infinite family of minimal binary linear codes, which do not satisfy Ashikhmin-Barg condition by using mathematical objects called simplicial complexes and generic construction
\begin{equation}\label{eq1.1}
\mathscr{C}_{f}=\{(\alpha f({\bf x})+{\bf w}\cdot {\bf x})_{\bf x\in \mathbb{F}_{p}^{n*}}\ | \ \alpha\in \mathbb{F}_{p}, \ {\bf w}\in \mathbb{F}_{p}^{n}\}
\end{equation}

The necessary and sufficient condition for binary linear code to be minimal was given by Ding, Heng, and Zhou \cite{9} in 2018, and they constructed three classes of minimal binary linear codes. Heng, Z., Ding, C., \& Zhou, Z. \cite{3}, derived necessary and sufficient  condition for ternary linear codes to be minimal and constructed a class of minimal linear codes, which violating Ashikhmin-Barg condition   from Lloyd polynomial. Also several researcher, construct minimal linear codes using different approaches \cite{5},\cite{6},\cite{7}      
\par \par In this paper, using the function defined in \cite{9}, we construct a class of minimal linear codes over $\mathbb{F}_{3}$, which satisfies $\frac{\text{wt}_{min}}{\text{wt}_{max}}\leq \frac{1}{3}<\frac{2}{3}$. Also we define new ternary function and construct minimal linear codes over $\mathbb{F}_{3}$, which violates Ashikhmin-Barg.
\par This paper is organised as follows: In Section 2, we give some basic definitions and results, which are prerequisites. Section 4 contains class of minimal linear codes over $\mathbb{F}_{3}$, using ternary functions. In Section 5, we obtain a class of minimal linear codes, which violates Ashikhmin-Barg condition using characteristic functions.
\section{Preliminaries}
In this section, we provide some definitions and results that are necessary prior knowledge for understanding this article. We denote a finite field with  $p$ elements by $\mathbb{F}_{p}$. A $[n,k,d]$ linear code over $\mathbb{F}_{3}$ is a $k$-dimensional subspace of $\mathbb{F}_{3}^{n}$, where $d$ is the minimum Hamming distance. Let $A_{i}$ denote the count of the codewords having weight $i$ in $\mathscr{C}$. The weight enumerator of a code $\mathscr{C}$ is defined as $1+A_{1}z+A_{2}z+\cdots + A_{n}z^{n}$, and the sequence $(1,A_{1},A_{2},\ldots, A_{n})$ is called the weight distribution of a code $\mathscr{C}$. If there are $t$ non-zero $A_{i}$, then $\mathscr{C}$ is called the $t$-weight code.
\begin{definition}\cite{10}
The $support$ $(\text{Supp}({\bf c}))$ of a codeword ${\bf c}$  is the set of the coordinates at which ${c}$  is non-zero. i.e.
\begin{equation*}
	\text{Supp}({\bf c})=\{ i \ | \ c_{i}\neq 0, \ 1\leq i\leq n\}
\end{equation*}
\end{definition}
Note that, $\text{wt}({\bf c})=|\text{Supp}({\bf c})|$, where $\text{wt}({\bf x})$ is the Hamming weight of ${\bf c}$. 
\begin{definition}
For ${\bf w}_{1},{\bf w}_{2}\in \mathbb{F}_{p}^{n}$, if $\text{Supp}({\bf w}_{1})\subseteq \text{Supp}({\bf w}_{2})$, then we say ${\bf w}_{2}$ covers ${\bf w}_{1}$ and denoted by ${\bf w}_{2}\preceq {\bf w}_{1}$.
\end{definition}
\begin{definition}
A codeword ${\bf w}$ is minimal if ${\bf w}$ covers only the codeword of the form $\alpha {\bf w}$, where $\alpha\in \mathbb{F}_{p}^{*}$. A code $\mathscr{C}$ is called minimal if every codeword in $\mathscr{C}$ is minimal.
\end{definition}
 \begin{definition}
 A ternary function $f$ is a function from $\mathbb{F}_{3}^{n}$ to $\mathbb{F}_{3}$. The Walsh transform for a ternary function is given by
 \begin{equation*}
 	\widehat{f}({\bf w})=\sum_{{\bf x}\in\mathbb{F}_{3}^{n}}\zeta_{3}^{f({\bf x})-{\bf w}\cdot{\bf x}}
 \end{equation*}
 where, $\zeta_{3}$ is a cube root of unity, ${\bf w} \in \mathbb{F}_{3}^{n}$ and ${\bf w}\cdot{\bf x}$ is the standard inner product of ${\bf w}$ and ${\bf x}$.
 \end{definition}
\begin{lemma}\cite{3}\label{1.1}
	For any ${\bf w}_{1},{\bf w}_{2}\in \mathbb{F}_{q}^{n}$, ${\bf w}_{1}\preceq{\bf w}_{2}$ if and only if 
	\begin{equation*}
	\sum_{{ c\in \mathbb{F}_{q}^{*}}}\text{wt}({\bf w }_{2}+c{\bf w}_{1})=(q-1)\text{wt}({\bf w}_{2})-\text{wt}({\bf w}_{1})
	\end{equation*}
\end{lemma}
\begin{theorem}\cite{3}\label{1.2}
Assume that $f({\bf x})\neq{\bf w}\cdot{\bf x}$ for any ${\bf w}\in \mathbb{F}_{3}^{n}$. The linear code $\mathscr{C}_{f}$ in \ref{eq1.1} has length $3^{n}-1$ and dimension $n+1$. In addition, the weight distribution of $\mathscr{C}_{f}$ is given by the following multiset union:
\begin{align*}
	&\left\{\ 2\left(3^{n-1}-\frac{\text{Re}(\widehat{f}({\bf w}))}{3} \right)\ | \ \alpha=-1, {\bf w}\in \mathbb{F}_{3}^{n}\right\}\cup\\
	& \left\{\ 2\left(3^{n-1}-\frac{\text{Re}(\widehat{f}({-\bf w}))}{3} \right)\ | \ \alpha=1, {\bf w}\in \mathbb{F}_{3}^{n}\right\}\cup\{3^{n}-3^{n-1}\ | \ \alpha=0, {\bf w}\in \mathbb{F}_{3}^{n}\}
\end{align*}
\end{theorem}
\begin{theorem}\cite{3}\label{1.3}
	Let $\mathscr{C}_{f}$ be the ternary code in \ref{eq1.1}. Assume that $f({\bf x})\neq{\bf w}\cdot{\bf x}$ for any ${\bf w}\in \mathbb{F}_{3}^{n}$. Then $\mathscr{C}_{f}$ is a minimal $[3^{n}-1,n+1]$ code if and only if
	\begin{align*}
		\text{Re}(\widehat{f}({\bf w}_{1}))+\text{Re}(\widehat{f}({\bf w}_{2}))+\text{Re}(\widehat{f}({\bf w}_{3}))&\neq 3^{n}
	\end{align*}
		\text{and}
		\begin{align*}
		\text{Re}(\widehat{f}({\bf w}_{1}))+\text{Re}(\widehat{f}({\bf w}_{2}))-2\text{Re}(\widehat{f}({\bf w}_{3}))&\neq 3^{n}\\
	\end{align*}
for all pairwise distinct ${\bf w}_{1}, {\bf w}_{2}$ and ${\bf w}_{3}$ with ${\bf w}_{1}+{\bf w}_{2}+{\bf w}_{3}={\bf 0}$.
\end{theorem}

\section{Construction of minimal linear codes over $\mathbb{F}_{3}$ from ternary functions}
In this section, first we define ternary functions using partial spread and using the ternary functions we construct minimal linear codes over $\mathbb{F}_{3}$, which violates Ashikhmin-Barg condition.
Throughout this paper, we assume that $n$ be a positive even integer and $t=\frac{n}{2}$. A partial spread of order $\nu$ in $\mathbb{F}_{3}^{n}$ is a collection of $\nu$ disjoint $t$-dimensional subspaces  $W_{1},W_{2},\ldots, W_{\nu}$ of $\mathbb{F}_{3}^{n}$. Also, observe that $\nu\leq 3^{t}+1$.\\
Let $f_{ij}:\mathbb{F}_{3}^{n}\rightarrow\mathbb{F}_{3}$ be ternary function given by 
\begin{equation*}
	f_{ij}({\bf x})=\begin{cases}
		1, & \text{ if } {\bf x}\in W_{i}\setminus\{0\}\\
		2, & \text{ if } {\bf x}\in W_{j}\setminus\{0\}\\
		0, & \text{Otherwise}
	\end{cases}
\end{equation*}
where, $i\neq j$ and $1\leq i,j\leq 3^{t}+1$. Note that $2f_{ij}=f_{ji}$.\\
For $1\leq s\leq \frac{3^{t}+1}{2}$ and $\mathcal{A}=\{i_{1},i_{2},\ldots,i_{s},i_{s+1},\ldots, i_{2s} \}$, we define ternary function $f$ as follows
\begin{equation}\label{eq4.1}
	f({\bf x})=\sum_{k=1}^{s}f_{i_{k}i_{s+k}}({\bf x})
\end{equation}
Using the function $f$ defined above, we will construct minimal linear code $\mathscr{C}_{f}$ over $\mathbb{F}_{3}$. For construction of minimal linear code $\mathscr{C}_{f}$, we need the values of $\text{Re}(\widehat{f}({\bf x}))$. The following lemma help us to find these values.
\begin{theorem}\label{5.1}
	Let $\mathbb{F}_{p}^{n}$ be an $n$-dimensional vector space over a finite field $\mathbb{F}_{p}$ and $W$ be a $m$-dimensional subspace of $\mathbb{F}_{p}^{n}$. If ${\bf y}\notin W^{\perp}$, then  $|\mathcal{S}_{k}|=p^{m-1}$, for all $k\in \mathbb{F}_{p}$, where $\mathcal{S}_{k}=\{{\bf x}\in W \ | \ {\bf y}\cdot {\bf x}=k \}$.
	\begin{proof}
		We prove this theorem by mathematical induction on $m$.\\
		If $m=1$, then we have $W=\text{span}(\{{\bf w}\})$. Since ${\bf y}\notin W^{\perp}$,
		we have ${\bf y}\cdot{\bf w}\neq0$, which implies ${\bf y}\cdot {\bf w}=k$, for some $k\in \mathbb{F}^{*}_{p}$. From this we get ${\bf x}\cdot\alpha{\bf w}=\alpha k $, for all $\alpha\in \mathbb{F}_{p}$. Hence  the result is true for $m=1$.\\
		Assume that result is true for $m-1$ and we prove the result  for $m$.\\
		 Let $W$ be a $m$-dimensional subspace of $\mathbb{F}_{p}^{n}$. Then $W=\text{span}(\{{\bf w_{1}}, {\bf w_{2}},\ldots, {\bf w_{m}} \})$. Since ${\bf y}\notin W^{\perp}$, we have ${\bf y}\cdot {\bf w}_{i}\neq 0$, for some $1\leq i\leq m$. Without loss of generality, consider the case ${\bf y}\cdot {\bf w}_{i}\neq 0$, for some $1\leq i\leq m-1 $ and ${\bf y}\cdot {\bf w}_{m}=k$, for some $k\in \mathbb{F}_{p}$. Now $W'=\text{span}(\{{\bf w}_{1},{\bf w}_{2},\ldots, {\bf w}_{m-1}\})$ is $(m-1)$-dimensional subspace of $\mathbb{F}_{p}^{n}$. Hence, the result is true for $W'$, from this we have, $\mathcal{S}'_{0}, \mathcal{S}'_{1}, \ldots, \mathcal{S}'_{p-1}$ be a collection of subsets of $W'$, with $|\mathcal{S}'_{i}|=p^{m-2}$, for all $0\leq i\leq p-1$ such that ${\bf y}\cdot{\bf x}=i$, for all ${\bf x}\in \mathcal{S}'_{i}$ and $0\leq i\leq p-1$. Note that $\mathcal{S}'_{0},\mathcal{S}'_{1},\ldots, \mathcal{S}'_{p-1}$ is also subsets of $W$. Now, consider the sets ${\bf w}_{m}+\mathcal{S}'_{0},{\bf w}_{m}+\mathcal{S}'_{1},\ldots, {\bf w}_{m}+\mathcal{S}'_{p-1}$ be a collection of subsets of $W$, with $|{\bf w}_{m}+\mathcal{S}'_{i}|=p^{m-2}$ and ${\bf y}\cdot({\bf w}_{m}+{\bf x})=k+i$, for all $0\leq i\leq p-1$ and for all ${\bf x}\in \mathcal{S}'_{i}$. Also, observe that $\mathcal{S}'_{i}\neq {\bf w}_{m}+\mathcal{S}'_{j}$, for $0\leq i,j\leq p-1$. Similarly, $2{\bf w}_{m}+\mathcal{S}'_{0},2{\bf w}_{m}+\mathcal{S}'_{1},\ldots, 2{\bf w}_{m}+\mathcal{S}'_{p-1}$ be a collection of subsets of $W$, with $|2{\bf w}_{m}+\mathcal{S}'_{i}|=p^{m-2}$ and ${\bf y}\cdot(2{\bf w}_{m}+{\bf x})=2k+i$, for all $0\leq i\leq p-1$ and for all ${\bf x}\in \mathcal{S}'_{i}$. Continuing in this manner, we get $\mathcal{S}_{0},\mathcal{S}_{1},\ldots, \mathcal{S}_{p-1}$ be a subset of $W$, with $|\mathcal{S}_{i}|=p^{m-1}$ such that ${\bf y}\cdot{\bf x}=i$, for ${\bf x}\in \mathcal{S}_{i}$. Hence, the result.
	\end{proof}
\end{theorem}  
\begin{remark}\label{remark}
	Note that, if $W_{1}$ and $W_{2}$ are disjoint subspace of a finite dimensional vector space $V$ over a field $\mathbb{F}_{p}$ such that
	\begin{equation*}
		\text{dim}(W_{1})+\text{dim}(W_{2})=\text{dim}(V)
	\end{equation*} then $W^{\perp}_{1}$ and $W^{\perp}_{2}$ are also disjoint subspaces of $V$.
\end{remark}
 Now following theorem give the weight distribution of $\mathscr{C}_{f}$ and the values of Walsh transform.
\begin{theorem}\label{3.1}
	Let $f$ be a ternary function defined in \ref{eq4.1}. Then $\mathscr{C}_{f}$ is a $[3^{n}-1,n+1]$ linear code over $\mathbb{F}_{3}$ with weight distribution given in Table \ref{table2}.	
		\begin{table}[ht]
			\centering
		\begin{tabular}{|c|c|}
			\hline
			Weight of ${\bf w}$ & Multiplicity $A_{{\bf w}}$\\
			\hline
			0 & 1\\
			$2s(3^{t}-1)$& 2\\
			$3^{n}-3^{n-1}$ & $3^{n}-1$\\
			$3^{n}-3^{n-1}-2s$ & $4s(3^{t}-1)$\\
			$3^{n}-3^{n-1}+3^{t}-2s$ & $2(3^{t}+1-2s)(3^{t}-1)$\\
			\hline
		\end{tabular}
		\caption{Weight Distribution of $\mathscr{C}_{f}\label{table2}$ }
	\end{table}
	\begin{proof}
		Note that $f({\bf x})\neq {\bf v}\cdot{\bf x}$, for all ${\bf v}\in \mathbb{F}_{3}^{n}$. Hence, from Theorem \ref{1.2}, we have $\mathscr{C}_{f}$ is a $[3^{n}-1,n+1]$ linear code over $\mathbb{F}_{3}$. Now, we find the weight distribution of the linear code $\mathscr{C}_{f}$. For any ${\bf w}\in \mathbb{F}_{3}^{n}$, we have three cases
		\begin{enumerate}[(I)]
			\item ${\bf w}={\bf 0}$
			\item ${\bf w}\notin W_{i}^{\perp}$, for all $i\in \mathcal{A}$.
			\item ${\bf w}\in W_{i}^{\perp}\setminus\{{\bf 0}\}$, for some $i\in \mathcal{A}$.
		\end{enumerate}
{\bf Case I:} If ${\bf w}=0$, then from the definition of $f_{ij}$ and $f$, we have
		\begin{align*}
			\widehat{f}({\bf 0})&=\sum_{{\bf x}\in \mathbb{F}_{3}^{n}}\zeta_{3}^{\sum_{k=1}^{s}f_{i_{k}i_{s+k}}({\bf x})-{\bf 0}\cdot {\bf x}}\\
			&= 1+\sum_{{\bf x}\in W_{i_{1}}\setminus\{0\}}\zeta_{3}^{1}+\sum_{{\bf x}\in W_{i_{s+1}}\setminus\{0\}}\zeta_{3}^{2}+\cdots +\sum_{{\bf x}\in W_{i_{s}}\setminus\{0\}}\zeta_{3}^{1}
			+\sum_{{\bf x}\in W_{i_{2s}}\setminus\{0\}}\zeta_{3}^{2}+\sum_{{\bf x}\notin \bigcup_{k=1}^{2s}W_{i_{k}}}\zeta_{3}^{0}
\end{align*}
Using Theorem \ref{5.1}, we get
\begin{align*}
			\widehat{f}({\bf 0}) &=1+\left(3^{t}-1\right)\zeta_{3}+\left(3^{t}-1\right)\zeta_{3}^{2}+\cdots
			+\left(3^{t}-1\right)\zeta_{3}+\left(3^{t-1}-1\right)\zeta_{3}^{2}+(3^{t}+1-2s)\left(3^{t}-1\right)\\
			&= 1+s(3^{t}-1)(\zeta_{3}+\zeta_{3}^{2})+(3^{t}+1-2s)\left(3^{t}-1\right)\\
			& =3^{n}-3s(3^{t}-1)
		\end{align*}
	{\bf Case II:}	If ${\bf w}\notin W_{i}^{\perp}$, for all $i\in \mathcal{A}$, then ${\bf w}\in W^{\perp}_{j}$, for some $j\notin\mathcal{A}$. Without loss of generality assume that ${\bf w}\in W^{\perp}_{i_{2s+1}}$. We have
		\begin{align*}
			\widehat{f}({\bf w})&=\sum_{{\bf x}\in\mathbb{F}_{3}^{n}}\zeta_{3}^{\sum_{k=1}^{s}f_{i_{k}i_{s+k}}({\bf x})-{\bf w}\cdot{\bf x}}\\
			&= 1+\sum_{{\bf x}\in W_{i_{1}}\setminus\{0\}}\zeta_{3}^{1-{\bf w}\cdot{\bf x}}+\sum_{{\bf x}\in W_{i_{s+1}}\setminus\{0\}}\zeta_{3}^{2-{\bf w}\cdot{\bf x}}\cdots +\sum_{{\bf x}\in W_{i_{s}}\setminus\{0\}}\zeta_{3}^{1-{\bf w}\cdot{\bf x}}\\
			&+\sum_{{\bf x}\in W_{i_{2s}}\setminus\{0\}}\zeta_{3}^{2-{\bf w}\cdot{\bf x}}+\sum_{{\bf x}\in W_{i_{2s+1}}\setminus\{0\}}\zeta_{3}^{0}+\sum_{{\bf x}\notin \bigcup_{k=1}^{2s+1}W_{i_{k}}}\zeta_{3}^{-{\bf w}\cdot{\bf x}}
\end{align*}
From Theorem \ref{5.1} and Remark \ref{remark}, we get
\begin{align*}
		\widehat{f}({\bf w})& =1+\left(3^{t-1}\zeta_{3}^{2}+(3^{t-1}-1)\zeta_{3}+3^{t-1}\right)+\left((3^{t-1}-1)\zeta_{3}^{2}+3^{t-1}\zeta_{3}+3^{t-1}\right)+\cdots\\
			&+\left(3^{t-1}\zeta_{3}^{2}+(3^{t-1}-1)\zeta_{3}+3^{t-1}\right)+\left((3^{t-1}-1)\zeta_{3}^{2}+3^{t-1}\zeta_{3}+3^{t-1}\right)\\
			&+(3^{t}-1)+(3^{t}+1-(2s+1))\left(3^{t-1}\zeta_{3}^{2}+3^{t-1}\zeta_{3}+(3^{t-1}-1)\right)\\
			&= 1-\zeta_{3}-\zeta_{3}^{2}-\cdots -\zeta_{3}-\zeta_{3}^{2}+(3^{t}-1)+(3^{t}+1-(2s+1))\left(3^{t-1}\zeta_{3}^{2}+3^{t-1}\zeta_{3}+(3^{t-1}-1)\right)\\
			& =3s
		\end{align*}
	{\bf Case III:} Let ${\bf w}\in W^{\perp}_{i}\setminus\{{\bf 0}\}$ for some $i\in \mathcal{A}$.	Suppose ${\bf w}\in W^{\perp}_{i}\setminus\{{\bf 0}\}$, for some $i\in \{i_{1},i_{2},\ldots, i_{s}\}$. Without loss of generality assume that ${\bf w}\in W^{\perp}_{i_{1}}\setminus\{{\bf 0}\}$.\\
	We have
		\begin{align*}
			\widehat{f}({\bf w})&=\sum_{{\bf x}\in\mathbb{F}_{3}^{n}}\zeta_{3}^{\sum_{k=1}^{s}f_{i_{k}i_{s+k}}({\bf x})-{\bf w}\cdot{\bf x}}\\
			&= 1+\sum_{{\bf x}\in W_{i_{1}}\setminus\{0\}}\zeta_{3}^{1}+\sum_{{\bf x}\in W_{i_{s+1}}\setminus\{0\}}\zeta_{3}^{2-{\bf w}\cdot{\bf x}}\cdots +\sum_{{\bf x}\in W_{i_{s}}\setminus\{0\}}\zeta_{3}^{1-{\bf w}\cdot{\bf x}}\\
			&+\sum_{{\bf x}\in W_{i_{2s}}\setminus\{0\}}\zeta_{3}^{2-{\bf w}\cdot{\bf x}}+\sum_{{\bf x}\notin \bigcup_{k=1}^{2s}W_{i_{k}}}\zeta_{3}^{1-{\bf w}\cdot{\bf x}}\\
			& =1+(3^{t}-1)\zeta_{3}+\left((3^{t-1}-1)\zeta_{3}^{2}+3^{t-1}\zeta_{3}+3^{t-1}\right)+\cdots\\
			&+\left(3^{t-1}\zeta_{3}^{2}+(3^{t-1}-1)\zeta_{3}+3^{t-1}\right)+\left((3^{t-1}-1)\zeta_{3}^{2}+3^{t-1}\zeta_{3}+3^{t-1}\right)\\
			&+(3^{t}+1-2s)\left(3^{t-1}\zeta_{3}^{2}+3^{t-1}\zeta_{3}+(3^{t-1}-1)\right)\\
			& =3^{t}(\zeta_{3}^{2}-1)+3s
		\end{align*}
		Similarly, if ${\bf w}\in W^{\perp}_{i}$, for some $i\in\{i_{s+1},i_{s+2},\ldots ,i_{2s}\}$,  then we get,\\
		$\widehat{f}({\bf w})=3^{t}(\zeta_{3}-1)+3s$. Hence, we get 
		\begin{equation*}
			\text{Re}(\widehat{f}({\bf w}))=\begin{cases}
				3^{n}-3s(3^{t}-1), & \text{ if } {\bf w}={\bf 0}\\
				3s, & \text{ if } {\bf w}\notin W^{\perp}_{i}, \text{ for all } i\in \mathcal{A}\\
				-\frac{3^{t+1}}{2}+3s, &\text{ if } {\bf w}\in W^{\perp}_{i}\setminus\{0\}, \text{ for some } i\in \mathcal{A}
			\end{cases}
		\end{equation*}
		The weight distribution of $\mathscr{C}_{f}$ follows from Theorem \ref{1.2}.
	\end{proof}
\end{theorem}
We now prove the following two lemmas, which are crucial in proving $\mathscr{C}_{f}$ is minimal.
\begin{lemma}\label{3.2}
	Let $n\geq 6$ and  $f$ be a ternary function defined in \ref{eq4.1}. If $s\neq \frac{3^{t}+1}{2}$, then
	\begin{equation*}
		\text{Re}(\widehat{f}({\bf w}_{1}))+\text{Re}(\widehat{f}({\bf w}_{2}))+\text{Re}(\widehat{f}({\bf w}_{3}))\neq 3^{n}
	\end{equation*}
	for all pairwise distinct ${\bf w}_{1},{\bf w}_{2},{\bf w}_{3}\in \mathbb{F}_{3}^{n}$ such that ${\bf w}_{1}+{\bf w}_{2}+{\bf w}_{3}={\bf 0}$.
	\begin{proof}
		For different choices of ${\bf w}_{1}, {\bf w}_{2}$ and ${\bf w}_{3}$, we have the following cases:\\
		{\bf Case I:} For ${\bf w}_{i}={\bf 0}$, for some $i$. Without loss of generality assume that ${\bf w}_{1}={\bf 0}$, then ${\bf w}_{2}=-{\bf w}_{3}$. Now, consider the following subcases:\\
		(a) If ${\bf w}_{2}\notin W^{\perp}_{i}$, for all $i\in \mathcal{A}$, then 
		\begin{align*}
			\text{Re}(\widehat{f}({\bf 0}))+\text{Re}(\widehat{f}({\bf w}_{2}))+\text{Re}(\widehat{f}({\bf w}_{3}))= 3^{n}-3s(3^{t}-1)+3s+3s\neq 3^{n} \text{ as } s\neq 0
		\end{align*}
		(b) If ${\bf w}_{2}\in W^{\perp}_{i}$, for some $i\in \mathcal{A}$, then
		\begin{align*}
			&\text{Re}(\widehat{f}({\bf 0}))+\text{Re}(\widehat{f}({\bf w}_{2}))+\text{Re}(\widehat{f}({\bf w}_{3}))= 3^{n}-3s(3^{t}-1)-\frac{3^{t+1}}{2}+3s-\frac{3^{t+1}}{2}+3s\neq 3^{n}\\
			&\text{ as } s\neq 3^{t-1}(s+1)
		\end{align*}
		{\bf Case II:} If ${\bf w}_{i}\neq 0$, for all $i\in \{1,2,3\}$, then consider the following subcases:\\
		(a) If ${\bf w}_{1},{\bf w}_{2},{\bf w}_{3}\notin W^{\perp}_{i}$, for all $i\in \mathcal{A}$, then 
		\begin{align*}
			&\text{Re}(\widehat{f}({\bf w}_{1}))+\text{Re}(\widehat{f}({\bf w}_{2}))+\text{Re}(\widehat{f}({\bf w}_{3}))= 3s+3s+3s= 9s\neq 3^{n}(\text{as } s\leq 3^{t}+1)
		\end{align*}
		(b) If ${\bf w}_{1},{\bf w}_{2},{\bf w}_{3}\in W^{\perp}_{i}$, for some $i\in \mathcal{A}$, then 
		\begin{align*}
			&\text{Re}(\widehat{f}({\bf w}_{1}))+\text{Re}(\widehat{f}({\bf w}_{2}))+\text{Re}(\widehat{f}({\bf w}_{3}))= \frac{-3^{t+1}}{2}+3s-\frac{3^{t+1}}{2}+3s-\frac{3^{t}+1}{2}+3s\neq 3^{n}\\
			&(\text{as }s\neq 3^{n-2}+\frac{3^{t}}{2})
		\end{align*}
		(c) If ${\bf w_{1}}\in W^{\perp}_{i}$, for some $i\in \mathcal{A}$ and ${\bf w_{2}},{\bf w_{3}}\notin W_{i} $, for all $i\in \mathcal{A}$, then
		\begin{align*}
			&\text{Re}(\widehat{f}({\bf w}_{1}))+\text{Re}(\widehat{f}({\bf w}_{2}))+\text{Re}(\widehat{f}({\bf w}_{3}))= -\frac{3^{t+1}}{2}+3s+3s+3s\neq 3^{n} (\text{as }s\leq 3^{t}+1)
		\end{align*}
		All other cases follows in similar way.\\
		Hence, we get 
		$\text{Re}(\widehat{f}({\bf w}_{1}))+\text{Re}(\widehat{f}({\bf w}_{2}))+\text{Re}(\widehat{f}({\bf w}_{3}))\neq 3^{n}$, for all pairwise distinct ${\bf w}_{1},{\bf w}_{2},{\bf w}_{3}\in \mathbb{F}_{3}^{n}$ such that ${\bf w}_{1}+{\bf w}_{2}+{\bf w}_{3}={\bf 0}$.
	\end{proof}
\end{lemma}
\begin{lemma}\label{3.3}
	Let $n\geq 6$ and  $f$ be a ternary function defined in \ref{eq4.1}. If $s\neq \frac{3^{t}+1}{2}$, then
	\begin{equation*}
		\text{Re}(\widehat{f}({\bf w}_{1}))+\text{Re}(\widehat{f}({\bf w}_{2}))-2\text{Re}(\widehat{f}({\bf w}_{3}))\neq 3^{n}
	\end{equation*}
	for all pairwise distinct ${\bf w}_{1},{\bf w}_{2},{\bf w}_{3}\in \mathbb{F}_{3}^{n}$ such that ${\bf w}_{1}+{\bf w}_{2}+{\bf w}_{3}={\bf 0}$.
	\begin{proof}
		For different choices of ${\bf w}_{1}, {\bf w}_{2}$ and ${\bf w}_{3}$, we have following cases:\\
		{\bf Case I:} If ${\bf w}_{1}={\bf 0}$, then ${\bf w}_{2}=-{\bf w}_{3}$. Now consider the following subcases:\\
		(a) If ${\bf w}_{2}\notin W^{\perp}_{i}$, for all $i\in \mathcal{A}$, then
		\begin{align*}
			&\text{Re}(\widehat{f}({\bf 0}))+\text{Re}(\widehat{f}({\bf w}_{2}))-2\text{Re}(\widehat{f}({\bf w}_{3}))= 3^{n}-3s(3^{t}-1)+3s-2(3s)\neq 3^{n}(\text{ as }s\neq 0)
		\end{align*}
		(b) If ${\bf w}_{2}\in W^{\perp}_{i}\setminus\{{\bf 0}\}$, for some $i\in \mathcal{A}$, then
		\begin{align*}
			&\text{Re}(\widehat{f}({\bf 0}))+\text{Re}(\widehat{f}({\bf w}_{2}))-2\text{Re}(\widehat{f}({\bf w}_{3}))= 3^{n}-3s(3^{t}-1)-\frac{3^{t+1}}{2}+3s-2\left(-\frac{3^{t+1}}{2}+3s\right)\neq 3^{n}\\
			&(\text{ as }s\neq \frac{1}{2})
		\end{align*}
		{\bf Case II:} If ${\bf w}_{3}={\bf 0}$, then ${\bf w}_{1}=-{\bf w}_{2}$. Now consider the following subcases:\\
		(a) If ${\bf w}_{1}\notin W^{\perp}_{i}$, for all $i\in \mathcal{A}$, then
		\begin{align*}
			&\text{Re}(\widehat{f}({\bf w}_{1}))+\text{Re}(\widehat{f}({\bf w}_{2}))-2\text{Re}(\widehat{f}({\bf 0}))= 3s+3s-2\left(3^{n}-3s(3^{t}-1)\right)\neq 3^{n} (\text{ as }s\neq \frac{3^{t}}{2})
		\end{align*}
		(b) If ${\bf w}_{1}\in W^{\perp}_{i}\setminus\{{\bf 0}\}$, for some $i\in \mathcal{A}$, then
		\begin{align*}
			&\text{Re}(\widehat{f}({\bf w}_{1}))+\text{Re}(\widehat{f}({\bf w}_{2}))-2\text{Re}(\widehat{f}({\bf 0}))= -\frac{3^{t+1}}{2}+3s-\frac{3^{t+1}}{2}+3s-2\left(3^{n}-3s(3^{t}-1)\right)\neq 3^{n}\\
			&(\text{as }s\neq \frac{3^{t}+1}{2})
		\end{align*}
		{\bf Case III:} If ${\bf w}_{i}\neq{\bf 0}$, for $i\in\{1,2,3\}$ then consider the following subcases:\\
		(a) If ${\bf w}_{1},{\bf w}_{2},{\bf w}_{3}\notin W^{\perp}_{i}$, for all $i\in \mathcal{A}$, then 
		\begin{align*}
			&\text{Re}(\widehat{f}({\bf w}_{1}))+\text{Re}(\widehat{f}({\bf w}_{2}))-2\text{Re}(\widehat{f}({\bf w}_{3}))= 3s+3s-2(3s)=0\neq 3^{n} 
		\end{align*}
		(b) If ${\bf w}_{1},{\bf w}_{2}\notin W^{\perp}_{i}$, for all $i\in \mathcal{A}$ and ${\bf w}_{3}\in W^{\perp}_{i}\setminus\{{\bf 0}\}$, for some $i\in \mathcal{A}$, then
		\begin{align*}
			&\text{Re}(\widehat{f}({\bf w}_{1}))+\text{Re}(\widehat{f}({\bf w}_{2}))-2\text{Re}(\widehat{f}({\bf w}_{3}))\neq 3^{n}\Leftrightarrow 3s+3s-2\left(-\frac{3^{t+1}}{2}+3s\right)=3^{t+1}\neq 3^{n}
		\end{align*}
		(c) If ${\bf w}_{1}\notin W^{\perp}_{i}$, for all $i\in \mathcal{A}$ and ${\bf w}_{2},{\bf w}_{3}\in W^{\perp}_{i}$, for some $i\in \mathcal{A}$, then
		\begin{align*}
			&\text{Re}(\widehat{f}({\bf w}_{1}))+\text{Re}(\widehat{f}({\bf w}_{2}))-2\text{Re}(\widehat{f}({\bf w}_{3}))= 3s-\frac{3^{t+1}}{2}+3s-2\left(-\frac{3^{t+1}}{2}+3s\right)=\frac{3^{t+1}}{2}\neq 3^{n}
		\end{align*}
		All other cases follows in similar way.\\
		Hence, we get 
		$\text{Re}(\widehat{f}({\bf w}_{1}))+\text{Re}(\widehat{f}({\bf w}_{2}))-2\text{Re}(\widehat{f}({\bf w}_{3}))\neq 3^{n}$, for all pairwise distinct ${\bf w}_{1},{\bf w}_{2},{\bf w}_{3}\in \mathbb{F}_{3}^{n}$ such that ${\bf w}_{1}+{\bf w}_{2}+{\bf w}_{3}={\bf 0}$.
	\end{proof}
\end{lemma}
\begin{theorem}
	Let $n\geq 6$ and  $f$ be a ternary function defined above. If $s\neq \frac{3^{t}+1}{2}$, then $\mathscr{C}_{f}$ is a $[2^{n}-1,n+1]$ minimal linear code. Moreover, if $s\leq 3^{t-2}$, then $\frac{\text{wt}_{min}}{\text{wt}_{max}}\leq \frac{1}{3}<\frac{2}{3}$.
	\begin{proof}
		From Lemma  \ref{3.2} and Lemma \ref{3.3}, $\mathscr{C}_{f}$ satisfies the necessary and sufficient condition of the Theorem \ref{1.3}. Hence $\mathscr{C}_{f}$ is minimal. \\
		If $s\leq 3^{t-2}$, then $\text{wt}_{min}=2s(3^{t}-1)$ and $\text{wt}_{max}=3^{n}-3^{n-1}+3^{t}-2s$.\\
		Then
		\begin{align*}
			\frac{\text{wt}_{min}}{\text{wt}_{max}}&=\frac{2s(3^{t}-1)}{3^{n}-3^{n-1}+3^{t}-2s}\\
			& \leq \frac{2\cdot3^{t-2}(3^{t}-1)}{2\cdot3^{2t-1}+3^{t}-2\cdot3^{t-2}}\\
			&= \frac{2(3^{t}-1)}{2\cdot3^{t+1}+7}\leq \frac{1}{3}<\frac{2}{3}
		\end{align*}
		Hence, the result.
	\end{proof}
\end{theorem}
\section{Construction of minimal linear codes over $\mathbb{F}_{3}$ from characteristic functions}
 In this section, we construct minimal linear codes over $\mathbb{F}_{3}$, using the function define in \cite{9}. First we define characteristic functions using partial spread and using these functions, we construct minimal linear codes over $\mathbb{F}_{3}$. \\
Consider a partial spread of order $\nu$ in $\mathbb{F}_{3}^{n}$ consisting  of $\nu$ disjoint $t$-dimensional subspaces  $W_{1},W_{2},\ldots, W_{\nu}$ of $\mathbb{F}_{3}^{n}$.\\
Let $f_{i}:\mathbb{F}_{3}^{n}\rightarrow\mathbb{F}_{3}$ be a characteristic  function given by 
\begin{align*}
f_{i}({\bf x})=\begin{cases}
	1, & \text{ if }  {\bf x}\in W_{i}\setminus\{0\}\\
	0 & \text{Otherwise}
\end{cases}
\end{align*}
For $1\leq s\leq 3^{t}+1$ and $\mathcal{A}=\{i_{1},i_{2},\ldots, i_{s}\}$, define the characteristic function $f$ as 
\begin{equation}\label{eq3.1}
f({\bf x})=\sum_{i\in \mathcal{A}}f_{i}({\bf x})
\end{equation}
 We now give  the weight distribution of $\mathscr{C}_{f}$ and values of the Walsh transform.
\begin{theorem}\label{2.1}
	Let $f$ be a characteristic  function defined in \ref{eq3.1}. Then $\mathscr{C}_{f}$ is a $[3^{n}-1,n+1]$ linear code over $\mathbb{F}_{3}$ with weight distribution given in Table \ref{table1}.	
	\begin{table}[ht]
		\centering
	\begin{tabular}{|c|c|}
		\hline
		Weight $w$ & Multiplicity $A_{w}$\\
		\hline
		0 & 1\\
		$s(3^{t}-1)$& 2\\
		$3^{n}-3^{n-1}$ & $3^{n}-1$\\
		$3^{n}-3^{n-1}-s$ & $2s(3^{t}-1)$\\
		$3^{n}-3^{n-1}+3^{t}-s$ & $2(3^{t}+1-s)(3^{t}-1)$\\
		\hline
	\end{tabular}
	\caption{Weight distribution of $\mathscr{C}_{f}$\label{table1} }
	\end{table}

\begin{proof}
Note that $f({\bf x})\neq {\bf v}\cdot{\bf x}$, for all ${\bf v}\in \mathbb{F}_{3}^{n}$. Hence, from Theorem \ref{1.2}, we have $\mathscr{C}_{f}$ is a $[3^{n}-1,n+1]$ linear code over $\mathbb{F}_{3}$. For any ${\bf w}\in \mathbb{F}_{3}^{n}$, we have three cases 
\begin{enumerate}[(I)]
	\item ${\bf w}={\bf 0}$
	\item ${\bf w}\notin W_{i}^{\perp}$, for all $i\in \mathcal{A}$.
	\item ${\bf w}\in W_{i}^{\perp}\setminus \{{\bf 0}\}$, for some $i\in \mathcal{A}$.
\end{enumerate}
{\bf Case I:} If ${\bf w}={\bf 0}$, then we have
\begin{align}\label{eq3.2}
\widehat{f}({\bf 0})&=\sum_{{\bf x}\in \mathbb{F}_{3}}\zeta_{3}^{\sum_{i\in\mathcal{A}}f_{i}({\bf x})-{\bf 0}\cdot {\bf x}}\nonumber\\
&= 1+\sum_{{\bf x}\in W_{i_{1}}\setminus\{0\}}\zeta_{3}^{1}+\sum_{{\bf x}\in W_{i_{2}}\setminus\{0\}}\zeta_{3}^{1}+\cdots +\sum_{{\bf x}\in W_{i_{s}}\setminus\{0\}}\zeta_{3}^{1}
+\sum_{{\bf x}\notin \bigcup_{k=1}^{s}W_{i_{k}}}\zeta_{3}^{0}\nonumber\\
& =1+\left(3^{t}-1\right)\zeta_{3}+\cdots
+\left(3^{t}-1\right)\zeta_{3}+(3^{t}+1-s)\left(3^{t}-1\right)\nonumber\\
& =3^{n}-3^{t}s+s+s(3^{t}-1)\zeta_{3}
\end{align}
{\bf Case II:} If ${\bf w}\notin W_{i}^{\perp}$, for all $i\in \mathcal{A}$ then ${\bf w}\in W^{\perp}_{j}$, for some $j\notin\mathcal{A}$. Without loss of generality, assume that ${\bf x}\in W_{i_{s+1}}^{\perp}$. We have 
\begin{align}\label{eq3.3}
\widehat{f}({\bf w})&=\sum_{{\bf x}\in \mathbb{F}_{3}^{n}}\zeta_{3}^{\sum_{i\in \mathcal{A}}f_{i}({\bf x})-w\cdot {\bf x}}\nonumber\\
&=1+\sum_{{\bf x}\in W_{i_{1}}\setminus\{0\}}\zeta_{3}^{1-{\bf w}\cdot{\bf x}}+\cdots +\sum_{{\bf x}\in W_{i_{s}}\setminus\{0\}}\zeta_{3}^{1-{\bf w}\cdot{\bf x}}+\sum_{{\bf x}\in W_{i_{s+1}}\setminus\{0\}}\zeta_{3}^{0}+\sum_{{\bf x}\notin\bigcup_{k=1}^{s+1}W_{i_{k}}}\zeta_{3}^{-{\bf w}\cdot{\bf x}}\nonumber\\
&= \left(3^{t-1}+(3^{t-1}-1)\zeta_{3}+3^{t-1}\zeta_{3}^{2}\right)+\cdots +\left(3^{t-1}+(3^{t-1}-1)\zeta_{3}+3^{t-1}\zeta_{3}^{2}\right)\nonumber\\
&+ (3^{t}-1)+\left(3^{t}+1-(s+1)\right)\left((3^{t-1}-1)+3^{t-1}\zeta_{3}+
3^{t-1}\zeta_{3}^{2}\right)\nonumber\\
&= -s\zeta_{3}+s
\end{align}
{\bf Case III:} For ${\bf w}\in W_{i}^{\perp}\setminus\{{\bf 0}\}$, for some $i\in \{i_{1},i_{2},\ldots, i_{s}\}$. Without loss of generality assume that ${\bf w}\in W_{i_{1}}^{\perp}\setminus\{{\bf 0}\}$. We have
\begin{align}\label{eq3.4}
	\widehat{f}({\bf w})&=\sum_{{\bf x}\in \mathbb{F}_{3}^{n}}\zeta_{3}^{\sum_{i\in \mathcal{A}}f_{i}({\bf x})-w\cdot {\bf x}}\nonumber\\
	&=1+\sum_{{\bf x}\in W_{i_{1}}\setminus\{0\}}\zeta_{3}^{1}+\cdots +\sum_{{\bf x}\in W_{i_{s}}\setminus\{0\}}\zeta_{3}^{1-{\bf w}\cdot{\bf x}}+\sum_{{\bf x}\notin\bigcup_{k=1}^{s}W_{i_{k}}}\zeta_{3}^{-{\bf w}\cdot{\bf x}}\nonumber\\
	&= (3^{t}-1)\zeta_{3}+\cdots +\left(3^{t-1}+(3^{t-1}-1)\zeta_{3}+3^{t-1}\zeta_{3}^{2}\right)\nonumber\\
	&+\left(3^{t}+1-s\right)\left((3^{t-1}-1)+3^{t-1}\zeta_{3}+
	3^{t-1}\zeta_{3}^{2}\right)\nonumber\\
	&= (3^{t}-1)\zeta_{3}-(s-1)\zeta_{3}-(3^{t}+1-s)
\end{align}
Hence, from \ref{eq3.2},\ref{eq3.3} and \ref{eq3.4}, we have 
\begin{equation*}
	\text{Re}(\widehat{f}({\bf w}))=\begin{cases}
		3^{n}-\frac{3}{2}s(3^{t}-1), & \text{ if } {\bf w}={\bf 0}\\
		\frac{3}{2}s, & \text{ if } {\bf w}\notin W^{\perp}_{i}, \text{ for all } i\in \mathcal{A}\\
		-\frac{3^{t+1}}{2}+\frac{3}{2}s, &\text{ if } {\bf w}\in W^{\perp}_{i}\setminus\{0\}, \text{ for some } i\in \mathcal{A}
	\end{cases}
\end{equation*}
Note that, $\text{Re}(\widehat{f}({\bf w}))=\text{Re}(2\widehat{f}({\bf w}))$.\\
The weight distribution of $\mathscr{C}_{f}$ follow from Theorem \ref{1.2}
\end{proof}
\end{theorem}
Now, we prove the following two results, which help us to prove $\mathscr{C}_{f}$ is a minimal linear code.
\begin{lemma}\label{2.2}
	Let $n\geq 6$ and  $f$ be a characteristic function defined in \ref{eq3.1}. If $s\notin \{1,3^{t},3^{t}+1\}$, then
	\begin{equation*}
		\text{Re}(\widehat{f}({\bf w}_{1}))+\text{Re}(\widehat{f}({\bf w}_{2}))+\text{Re}(\widehat{f}({\bf w}_{3}))\neq 3^{n}
	\end{equation*}
	for all pairwise distinct ${\bf w}_{1},{\bf w}_{2},{\bf w}_{3}\in \mathbb{F}_{3}^{n}$ such that ${\bf w}_{1}+{\bf w}_{2}+{\bf w}_{3}={\bf 0}$.
	\begin{proof}
		For different choices of ${\bf w}_{1}, {\bf w}_{2}$ and ${\bf w}_{3}$, we have following subcases:\\
		{\bf Case I:} For ${\bf w}_{i}={\bf 0}$, for some $i$. Without loss of generality assume that ${\bf w}_{1}={\bf 0}$, then ${\bf w}_{2}=-{\bf w}_{3}$. Now consdier the following subcases:\\
		(a) If ${\bf w}_{2}\notin W^{\perp}_{i}$, for all $i\in \mathcal{A}$, then 
		\begin{align*}
			\text{Re}(\widehat{f}({\bf 0}))+\text{Re}(\widehat{f}({\bf w}_{2}))+\text{Re}(\widehat{f}({\bf w}_{3}))= 3^{n}-\frac{3}{2}s(3^{t}-1)+\frac{3}{2}s+\frac{3}{2}s\neq 3^{n} (\text{since }s\neq 0)
		\end{align*}
		(b) If ${\bf w}_{2}\in W^{\perp}_{i}\setminus\{{\bf 0}\}$, for some $i\in \mathcal{A}$, then
		\begin{align*}
			&\text{Re}(\widehat{f}({\bf 0}))+\text{Re}(\widehat{f}({\bf w}_{2}))+\text{Re}(\widehat{f}({\bf w}_{3}))= 3^{n}-\frac{3}{2}s(3^{t}-1)-\frac{3^{t+1}}{2}+\frac{3}{2}s-\frac{3^{t+1}}{2}+\frac{3}{2}s\neq 3^{n}\\
			& (\text{since } s\neq 3^{t-1}(s+2))
		\end{align*}
		{\bf Case II:} If ${\bf w}_{i}\neq 0$, for $i=1,2,3$, then consider the following subcases:\\
		(a) If ${\bf w}_{1},{\bf w}_{2},{\bf w}_{3}\notin W^{\perp}_{i}$, for all $i\in \mathcal{A}$, then 
		\begin{align*}
			&\text{Re}(\widehat{f}({\bf w}_{1}))+\text{Re}(\widehat{f}({\bf w}_{2}))+\text{Re}(\widehat{f}({\bf w}_{3}))= \frac{3}{2}s+\frac{3}{2}s+\frac{3}{2}s\neq 3^{n}=\frac{9}{2} s\neq 3^{n}
		\end{align*}
		(b) If ${\bf w}_{1},{\bf w}_{2},{\bf w}_{3}\in W^{\perp}_{i}\setminus\{{\bf 0}\}$, for some $i\in \mathcal{A}$, then 
		\begin{align*}
			&\text{Re}(\widehat{f}({\bf w}_{1}))+\text{Re}(\widehat{f}({\bf w}_{2}))+\text{Re}(\widehat{f}({\bf w}_{3}))= -\frac{3^{t+1}}{2}+\frac{3}{2}s-\frac{3^{t+1}}{2}+\frac{3}{2}s-\frac{3^{t+1}}{2}+\frac{3}{2}s\\
			&=-\frac{3^{t+2}}{2}+\frac{9}{2}s\neq 3^{n}(\text{since }s\leq 3^{t}+1)			
		\end{align*}
		(c) If ${\bf w_{1}}\in W^{\perp}_{i}\setminus\{{\bf 0}\}$, for some $i\in \mathcal{A}$ and ${\bf w_{2}},{\bf w_{3}}\notin W^{\perp}_{i} $, for all $i\in \mathcal{A}$, then
		\begin{align*}
			&\text{Re}(\widehat{f}({\bf w}_{1}))+\text{Re}(\widehat{f}({\bf w}_{2}))+\text{Re}(\widehat{f}({\bf w}_{3}))= -\frac{3^{t+1}}{2}+\frac{3}{2}s+\frac{3}{2}s+\frac{3}{2}s=-\frac{3^{t+1}}{2}+\frac{9}{2}s\neq 3^{n}\\
			& (\text{as }s\leq 3^{t}+1)
		\end{align*}
		All other cases follows in similar way.\\
		Hence, we get 
		$\text{Re}(\widehat{f}({\bf w}_{1}))+\text{Re}(\widehat{f}({\bf w}_{2}))+\text{Re}(\widehat{f}({\bf w}_{3}))\neq 3^{n}$, for all pairwise distinct ${\bf w}_{1},{\bf w}_{2},{\bf w}_{3}\in \mathbb{F}_{3}^{n}$ such that ${\bf w}_{1}+{\bf w}_{2}+{\bf w}_{3}={\bf 0}$.
	\end{proof}
\end{lemma}
\begin{lemma}\label{2.3}
	Let $n\geq 6$ and  $f$ be a characteristic function defined in \ref{eq3.1}. If $s\notin \{1,3^{t},3^{t}+1\}$, then
	\begin{equation*}
		\text{Re}(\widehat{f}({\bf w}_{1}))+\text{Re}(\widehat{f}({\bf w}_{2}))-2\text{Re}(\widehat{f}({\bf w}_{3}))\neq 3^{n}
	\end{equation*}
	for all pairwise distinct ${\bf w}_{1},{\bf w}_{2},{\bf w}_{3}\in \mathbb{F}_{3}^{n}$ such that ${\bf w}_{1}+{\bf w}_{2}+{\bf w}_{3}={\bf 0}$.
	\begin{proof}
		For different choices of ${\bf w}_{1}, {\bf w}_{2}$ and ${\bf w}_{3}$, we have following subcases:\\
		{\bf Case I:} If ${\bf w}_{1}={\bf 0}$, then ${\bf w}_{2}=-{\bf w}_{3}$. Now consider the following subcases:\\
		(a) If ${\bf w}_{2}\notin W^{\perp}_{i}$, for all $i\in \mathcal{A}$, then
		\begin{align*}
			&\text{Re}(\widehat{f}({\bf 0}))+\text{Re}(\widehat{f}({\bf w}_{2}))-2\text{Re}(\widehat{f}({\bf w}_{3}))=-\frac{3}{2}s(3^{t}-1)+\frac{3}{2}s-2\left(\frac{3}{2}s\right)\neq 3^{n}(\text{as }s\neq 0)
		\end{align*}
		(b) If ${\bf w}_{2}\in W^{\perp}_{i}\setminus\{{\bf 0}\}$, for some $i\in \mathcal{A}$, then
		\begin{align*}
			&\text{Re}(\widehat{f}({\bf 0}))+\text{Re}(\widehat{f}({\bf w}_{2}))-2\text{Re}(\widehat{f}({\bf w}_{3}))= 3^{n}-\frac{3}{2}s(3^{t}-1)-\frac{3^{t+1}}{2}+\frac{3}{2}s-2\left(-\frac{3^{t+1}}{2}+\frac{3}{2}s\right)\\
			&= 3^{n}-\frac{3^{t+1}}{2}(s-1)\neq 3^{n}(\text{since }s\neq 1)
		\end{align*}
		{\bf Case II:} If ${\bf w}_{3}={\bf 0}$, then ${\bf w}_{1}=-{\bf w}_{2}$. Now consider the following subcases:\\
		(a) If ${\bf w}_{1}\notin W^{\perp}_{i}$, for all $i\in \mathcal{A}$, then
		\begin{align*}
			&\text{Re}(\widehat{f}({\bf w}_{1}))+\text{Re}(\widehat{f}({\bf w}_{2}))-2\text{Re}(\widehat{f}({\bf 0}))= \frac{3}{2}s+\frac{3}{2}s-2\left(3^{n}-\frac{3}{2}s(3^{t}-1)\right)\\
			&=- 2\cdot3^{n}+3^{t+1}s\neq 3^{n} (\text{since }s\neq 3^{t})
		\end{align*}
		(b) If ${\bf w}_{1}\in W^{\perp}_{i}\setminus\{{\bf 0}\}$, for some $i\in \mathcal{A}$, then
		\begin{align*}
			&\text{Re}(\widehat{f}({\bf w}_{1}))+\text{Re}(\widehat{f}({\bf w}_{2}))-2\text{Re}(\widehat{f}({\bf 0}))= -\frac{3^{t+1}}{2}+\frac{3}{2}s-\frac{3^{t+1}}{2}+\frac{3}{2}s-2\left(3^{n}-\frac{3}{2}s(3^{t}-1)\right)\\
			&= -2\cdot3^{n}-3^{t+1}+3^{t+1}s\neq 3^{n} (\text{since }s\neq 3^{t}+1)
		\end{align*}
		{\bf Case III:} If ${\bf w}_{i}\neq{\bf 0}$, for all $i\in\{1,2,3\}$ then consider the following subcases:\\
		(a) If ${\bf w}_{1},{\bf w}_{2},{\bf w}_{3}\notin W^{\perp}_{i}$, for all $i\in \mathcal{A}$, then 
		\begin{align*}
			&\text{Re}(\widehat{f}({\bf w}_{1}))+\text{Re}(\widehat{f}({\bf w}_{2}))-2\text{Re}(\widehat{f}({\bf w}_{3}))= \frac{3}{2}s+\frac{3}{2}s-2\left(\frac{3}{2}s\right)=0\neq 3^{n}
		\end{align*}
		(b) If ${\bf w}_{1},{\bf w}_{2}\notin W^{\perp}_{i}$, for all $i\in \mathcal{A}$ and ${\bf w}_{3}\in W^{\perp}_{i}\setminus\{{\bf 0}\}$, for some $i\in \mathcal{A}$, then
		\begin{align*}
			&\text{Re}(\widehat{f}({\bf w}_{1}))+\text{Re}(\widehat{f}({\bf w}_{2}))-2\text{Re}(\widehat{f}({\bf w}_{3}))= \frac{3}{2}s+\frac{3}{2}s-2\left(-\frac{3^{t+1}}{2}+\frac{3}{2}s\right)=3^{t+1}\neq 3^{n}
		\end{align*}
		(c) If ${\bf w}_{1}\notin W^{\perp}_{i}$, for all $i\in \mathcal{A}$ and ${\bf w}_{2},{\bf w}_{3}\in W_{i}^{\perp}\setminus\{{\bf 0}\}$, for some $i\in \mathcal{A}$, then
		\begin{align*}
			&\text{Re}(\widehat{f}({\bf w}_{1}))+\text{Re}(\widehat{f}({\bf w}_{2}))-2\text{Re}(\widehat{f}({\bf w}_{3}))= \frac{3}{2}s-\frac{3^{t+1}}{2}+\frac{3}{2}s-2\left(-\frac{3^{t+1}}{2}+\frac{3}{2}s\right)= \frac{3^{t+1}}{2}\neq 3^{n}
		\end{align*}
		All other cases follows in similar way.\\
		Hence, we get 
		$\text{Re}(\widehat{f}({\bf w}_{1}))+\text{Re}(\widehat{f}({\bf w}_{2}))-2\text{Re}(\widehat{f}({\bf w}_{3}))\neq 3^{n}$, for all pairwise distinct ${\bf w}_{1},{\bf w}_{2},{\bf w}_{3}\in \mathbb{F}_{3}^{n}$ such that ${\bf w}_{1}+{\bf w}_{2}+{\bf w}_{3}={\bf 0}$.
	\end{proof}
\end{lemma}
Now, we provide a class of minimal linear codes violating Ashikhmin-Barg condition.
\begin{theorem}
	Let $n\geq 6$ and   $f$ be a characteristic function defined in \ref{eq3.1}. If $s\notin \{1,3^{t},3^{t}+1\}$, then $\mathscr{C}_{f}$ is a $[2^{n}-1,n+1]$ minimal linear code. Moreover, if $s\leq 3^{t-2}$, then $\frac{\text{wt}_{min}}{\text{wt}_{max}}\leq\frac{1}{3} <\frac{2}{3}$.
	\begin{proof}
		From Lemma \ref{2.2} and \ref{2.3}, $\mathscr{C}_{f}$ satisfies conditions in Theorem \ref{1.3}. Hence, $\mathscr{C}_{f}$ is minimal. \\
		If $s\leq 3^{t-2}$, then $\text{wt}_{min}=s(3^{t}-1)$ and $\text{wt}_{max}=3^{n}-3^{n-1}+3^{t}-s$.
			\begin{align*}
	\text{Therefore,}	\quad	\frac{\text{wt}_{min}}{\text{wt}_{max}}&=\frac{s(3^{t}-1)}{3^{n}-3^{n-1}+3^{t}-s}\\
			& \leq \frac{3^{t-2}(3^{t}-1)}{2\cdot3^{2t-1}+3^{t}-3^{t-2}}\\
			&= \frac{(3^{t}-1)}{2\cdot3^{t+1}+8}\leq \frac{1}{3}<\frac{2}{3}
		\end{align*}
		Hence, the result.
	\end{proof}
\end{theorem}

\section{Conclusion}
In this paper, we constructed two classes of ternary minimal linear codes of dimension $n+1$. One class is constructed using characteristic function over partial spread and another class is constructed using ternary function on partial spread. We also obtained weight distribution for these minimal linear codes. We further prove that a family of these constructed minimal linear codes violates Ashikhmin-Barg condition.  Using the functions defined in this paper, constructing minimal linear codes over  $\mathbb F_{3}$ of dimension greater than $n+1$  can be a further scope of the study.

\end{document}